\documentclass[12pt]{article}
\usepackage{epsfig,amsmath}
\def\beqn{\begin{eqnarray}}
\def\eeqn{\end{eqnarray}}
\def\vs{\vspace{0.3cm}}

\begin{document}
\pagestyle{empty}
\baselineskip=0.212in
\renewcommand{\theequation}{\arabic{section}.\arabic{equation}}
\renewcommand{\thefigure}{\arabic{figure}}
\renewcommand{\thetable}{\arabic{section}.\arabic{table}}

\begin{flushleft}
\large
{SAGA-HE-125-97
\hfill November 25, 1997}  \\
\end{flushleft}

\vspace{1.5cm}
 
\begin{center}
 
\Large{{\bf Numerical solution of $\bf Q^2$ evolution equation}} \\
\vspace{0.3cm}
\Large{{\bf for the transversity distribution $\bf\Delta_{_T} q$}} \\

\vspace{1.3cm}
 
\Large
{M. Hirai, S. Kumano, and M. Miyama $^*$ }         \\
 
\vspace{0.5cm}
 
\Large
{Department of Physics, Saga University, Saga 840, Japan} \\

\vspace{1.5cm}

\Large{ABSTRACT}
  
\end{center}

We investigate numerical solution of 
the Dokshitzer-Gribov-Lipatov-Altarelli-Parisi
(DGLAP) $Q^2$ evolution equation for
the transversity distribution $\Delta_{_T} q$
or the structure function $h_1$.
The leading-order (LO) and next-to-leading-order (NLO)
evolution equations are studied. 
The renormalization scheme is $MS$ or $\overline{MS}$ in the NLO case.
Dividing the variables $x$ and $Q^2$ into small steps, we solve
the integrodifferential equation by the Euler method in
the variable $Q^2$ and by the Simpson method in the variable $x$.
Numerical results indicate that accuracy is better than 1\%
in the region $10^{-5}<x<0.8$ if more than fifty $Q^2$ steps
and more than five hundred $x$ steps are taken.
We provide a FORTRAN program for the Q$^2$ evolution and devolution
of the transversity distribution $\Delta_{_T} q$ or $h_1$.
Using the program, we show the LO and NLO evolution results of 
the valence-quark distribution $\Delta_{_T} u_v + \Delta_{_T} d_v$,
the singlet distribution $\sum_i (\Delta_{_T} q_i + \Delta_{_T} \bar q_i)$,
and the flavor asymmetric distribution 
$\Delta_{_T} \bar u - \Delta_{_T} \bar d$.
They are also compared with the longitudinal evolution results.

\vspace{0.8cm}

\vfill
 
\noindent
{\rule{6.cm}{0.1mm}} \\
 
\vspace{-0.3cm}
\normalsize
\noindent
* Email: 96sm18@edu.cc.saga-u.ac.jp, 
          kumanos@cc.saga-u.ac.jp, and \\

\vspace{-0.6cm}
\noindent
96td25@edu.cc.saga-u.ac.jp. Information on their research is available \\

\vspace{-0.6cm}
\noindent
at http://www.cc.saga-u.ac.jp/saga-u/riko/physics/quantum1/structure.html. \\

\vspace{+0.1cm}
\hfill
{submitted for publication}

\normalsize


\vfill\eject
\pagestyle{plain}
\setcounter{page}{1}

\section*{{\bf Program Summary}}
\addcontentsline{toc}{section}{\protect\numberline{\S}{Program Summary}}

\ \\
\noindent
{\it Title of program:} H1EVOL

\vs
\noindent
{\it Computer:} AlphaServer 2100 4/200 (SUN-IPX); 
               {\it Installation:} 
                     The Research Center for Nuclear Physics in Osaka
                     (Saga University Computer Center)

\vs
\noindent
{\it Operating system:} OpenVMS V6.1 (SunOS 4.1.3)

\vs
\noindent
{\it Programming language used:} FORTRAN 77

\vs
\noindent
{\it Peripherals used:} Laser printer

\vs
\noindent
{\it No. of lines in distributed program:} 
1203

\vs
\noindent
{\it Keywords:} Polarized parton distribution, transversity distribution,
                chiral-odd structure function, 
                Q$^2$ evolution, numerical solution.

\vs
\noindent
{\it Nature of physical problem} 

\noindent
This program solves the DGLAP $Q^2$ evolution equation
with or without next-to-leading-order (NLO) $\alpha_s$ effects
for a transversely polarized parton distribution, so called
transversity distribution $\Delta_{_T} q$ or
chiral-odd structure function $h_1$.

\vs
\noindent
{\it Method of solution}

\noindent
The DGLAP integrodifferential equation is solved
by dividing the variables $x$ and $Q^2$ into very small steps.
The integrodifferential equation is solved step by step
by the Euler method (``brute-force" method)
in the variable $Q^2$ and by the Simpson method
in the variable $x$.

\vs
\noindent
{\it Restrictions of the program}

\noindent
This program is used for calculating the $Q^2$ evolution of
various transversity distributions $\Delta_{_T} q$ ($h_1$).
The $Q^2$ evolution equation is the DGLAP equation.
The double precision arithmetic is used.
The  minimal subtraction ($MS$) or 
the modified minimal subtraction ($\overline{MS}$) scheme is used 
in the NLO evolution. The NLO evolution is identical
in both schemes.
A user provides the initial distribution as a subroutine
or as a data file.
Examples are explained in Section \ref{INPUT}.
Then, the user inputs seventeen parameters in Section \ref{INPUT}.

\vs
\noindent
{\it Typical running time}

\noindent
Approximately forty seconds on AlphaServer 2100 4/200 
and two and a half minutes on SUN-IPX.

\vfill\eject
\section*{{\bf LONG WRITE-UP}}

\noindent
\section{{\bf Introduction}}\label{INTRO}
\setcounter{equation}{0}
\setcounter{figure}{0}
\setcounter{table}{0}

The European Muon Collaboration (EMC) shed light on
the proton spin structure by measuring the longitudinally polarized
structure function $g_1$. Its conclusion seemed startling: almost none
of the proton spin is carried by quarks.
Since then, many investigations have been done for understanding
the internal spin structure of the proton. 
However, we should wait for refined
experiments in order to determine each parton polarization,
in particular polarized sea-quark and gluon distributions 
\cite{RHIC-SPIN-J}.

In addition to the abovementioned studies on the longitudinal polarization,
it is interesting to test the spin structure in the transverse
polarization. There exists a leading-twist structure function which
will be measured in the transversely polarized Drell-Yan processes
\cite{RS,H1,DYEXP}. It is now named as the $h_1$ structure
function \cite{H1}, which measures the transversity distribution,
$\Delta_{_T} q$ or often denoted as $\delta q$.
It has a chiral-odd property so that it cannot be found
in inclusive deep inelastic electron scattering.
The $h_1$ structure function is expected to be measured
in the Relativistic Heavy Ion Collider (RHIC) Spin project
\cite{RHIC-SPIN} at Brookhaven National Laboratory.
The experimental results will provide us a good opportunity
to understand unexplored transverse spin physics at high energies.
Furthermore, semi-inclusive electron deep inelastic scattering
could also reveal the $h_1$ although there is a complication of
unknown fragmentation functions \cite{SEMI-H1}.
With these experimental possibilities in mind, theoretical physicists
should investigate detailed properties of $h_1$ \cite{H1PAPER}
as far as they can before the completion of the RHIC facility.

There are, in general, two variables in the structure functions: 
$Q^2$ and the momentum fraction $x$. 
In the transversity distribution measured by the Drell-Yan,
$Q^2$ is the dimuon-mass square: $Q^2=m_{\mu\mu}^2$.
The $x$ distribution is associated with nonperturbative physics,
which can be handled only in certain quark models at this stage 
because there is no experimental information on $h_1$.
The lattice QCD calculation is not still very reliable
as far as the $x$ distribution is concerned \cite{GHPRS3}.
On the other hand, the $Q^2$ dependence can be predicted in
perturbative Quantum Chromodynamics (QCD) unless $Q^2$ is small. 
This scaling violation is described by the
Dokshitzer-Gribov-Lipatov-Altarelli-Parisi (DGLAP) equation \cite{AP},
which is an integrodifferential equation.
The leading-order (LO) DGLAP equation for $h_1$ was derived
in Ref. \cite{AM} and its numerical analysis is discussed
in Ref. \cite{BCD}. 
In addition, the next-to-leading order (NLO) analysis for 
its anomalous dimensions and splitting function was completed
recently for the transversity distribution $\Delta_{_T} q$.
The NLO evolution of $\Delta_{_T} q$ was studied in the Feynman
gauge \cite{KM} and in the light-cone gauge \cite{V}.
It was also investigated in the Feynman gauge
in a subsequent paper \cite{HKK,COM}. 
The results made it possible to investigate
the details of NLO evolution effects in the structure function $h_1$
\cite{SV,BST,MSSV}. In particular, the NLO analysis is usually used
in the unpolarized structure functions and in the longitudinally
polarized structure function $g_1$. The NLO studies can be extended
now to the $h_1$ case. 
In this paper, we report our studies on the numerical solution.

Our group has been investigating the numerical solution 
in the unpolarized structure functions and also in 
the polarized structure function $g_1$.
So far, we employed the Laguerre-polynomial \cite{LAG} and
the brute-force \cite{BF} methods.
Other methods with the Mellin transformation are, for example,
discussed in Ref. \cite{BGR}.
The variables $Q^2$ and $x$ are divided into small steps, then the
integrodifferential equation is solved step by step in Ref. \cite{BF}.
Because the scaling violation is a small logarithmic
effect, the integration over $Q^2$ is accurately done
with a small number of steps. In this paper, we change
the brute-force integration over $x$ to the Simpson method
with expectation of better accuracy.
The transversity evolution is rather simple in the sense
that the gluon does not couple to the distribution $\Delta_{_T} q$
because of the chiral-odd nature. Therefore, the evolution equation
is a single integrodifferential equation without mixing with
the gluon distribution. It is the same form as the nonsinglet
evolution equation. This fact simplifies our numerical analysis.
The transversity evolution is completed by solving only
the ``nonsinglet-type" evolution equation.

The purpose of our study is to provide a useful $Q^2$ evolution
program for the distribution $\Delta_{_T} q$ (or $h_1$).
It should be very useful for theoretical and experimental researchers
who are involved in the RHIC-Spin type project or in semi-inclusive
lepton scattering one for finding $h_1$.
We explain the details of our studies in the following.
In Section \ref{DGLAP}, the transversity DGLAP $Q^2$ evolution equation
is explained. 
Then, our numerical-solution method
is described in Section \ref{SIMPSON}.
Input parameters and input distributions are discussed in 
Section \ref{INPUT}, and subroutines in the program are 
explained in Section \ref{H1EVOL}.
Numerical results and their comparison with longitudinally polarized
distributions are discussed in Section \ref{ANALYSIS}. 
Summary is given in Section \ref{SUMMARY}.
Explicit forms of splitting functions are listed in Appendices.

\vfill\eject
\section{{\bf $\bf Q^2$ evolution equation for the transversity \\
              distribution $\bf\Delta_{_T} q$}}\label{DGLAP}
\setcounter{equation}{0}
\setcounter{figure}{0}
\setcounter{table}{0}

The transversity distribution $\Delta_{_T} q_i$, which
is equivalently denoted as $h_1^i$ or $\delta q_i$, 
with the quark flavor $i$ can be measured, for example,
in the double transverse-spin asymmetry $A_{TT}$ in the Drell-Yan process
\cite{RS,H1,DYEXP}:
\begin{equation}
A_{TT}= \frac{\sin^2 \theta \, \cos 2 \phi}{1+\cos^2\theta}
\frac{\sum _i  e_i^2 \, \Delta_{_T} q_i (x_1) \, 
                          \Delta_{_T} \bar q_i (x_2)}
     {\sum _i  e_i^2 \, q_i (x_1) \, \bar q_i (x_2) }
\ .
\label{eqn:ATT}
\end{equation}
The factor $a_{_{TT}}=\sin^2 \theta \, \cos 2 \phi /(1+\cos^2\theta)$
is the parton's double asymmetry.
The angles $\theta$ and $\phi$ are the polar and azimuthal
angles of the lepton momentum with respect to the beam and proton
polarization directions respectively.
According to the above equation, the distribution combination 
$\Delta_{_T} q_i \, \Delta_{_T} \bar q_i$ weighted by the charge square
can be measured with information on
the unpolarized distributions $q_i$ and $\bar q_i$.
The spin asymmetry $A_{TT}$ will be measured in the RHIC-Spin project
\cite{RHIC-SPIN}, and the transversity distributions could be obtained
by Eq. (\ref{eqn:ATT}).

The distribution $\Delta_{_T} q$ is interpreted in a parton model:
it is the probability to find a quark with spin polarized along
the transverse spin of a polarized proton minus
the probability to find it polarized oppositely
($\Delta_{_T} q = q_{\uparrow}-q_{\downarrow}$).
The LO evolution equation for $\Delta_{_T} q$ was derived
in Ref. \cite{AM}, and the NLO evolution of $\Delta_{_T} q$
was recently obtained in Refs. \cite{KM,V,HKK}.
Because of the chiral-odd nature of the transversity distribution
$\Delta_{_T} q$ or the structure function $h_1$, the gluon does not
participate in the evolution equation. Therefore, the DGLAP equation
is a single integrodifferential equation:
\begin{equation}
\frac{\partial}{\partial\ln Q^2} \, \Delta_{_T} q^\pm (x,Q^2) \, = 
\frac{\alpha_s (Q^2)}{2\pi} \,  
\Delta_{_T} P_{q^\pm} (x) \otimes \Delta_{_T} q^\pm (x,Q^2) 
\ ,
\label{eqn:DGLAP1}
\end{equation}
where $\Delta_{_T} P_{q^\pm}$ is the splitting function for the
transversity distribution. 
The convolution $\otimes$ in Eq. (\ref{eqn:DGLAP1}) is defined by
\begin{equation}
f (x) \otimes g (x) = \int^{1}_{x} \frac{dz}{z} 
f\left(\frac{x}{z} \right) g(z) 
\ .
\label{eqn:conv}
\end{equation}
The notation $q^\pm$ in the splitting function indicates
the $\Delta_{_T} q^+ = \Delta_{_T} q + \Delta_{_T} \bar q$ or 
    $\Delta_{_T} q^- = \Delta_{_T} q - \Delta_{_T} \bar q$
distribution type.
The $\alpha _s (Q^2)$ is the running coupling constant. 
Both LO and NLO evolution calculations can be handled by 
Eq. (\ref{eqn:DGLAP1}); however, there are differences
between the LO and NLO splitting functions and 
also between the coupling constants. 
The LO and NLO expressions for $\Delta_{_T} P_{q^\pm}$ \cite{V}
and $\alpha _s$ are listed in Appendix A.
In the following numerical analysis,
we use the variable $t$ which is defined by
\begin{equation}
t \equiv \ln Q^2
\end{equation}
instead of $Q^2$.
Furthermore, the parton distribution and the splitting function multiplied
by $x$, 
\beqn
\widetilde f (x) = x f(x)
\ ,
\eeqn
are used throughout this paper because they
satisfy the same integrodifferential equation.
Then, Eq. (\ref{eqn:DGLAP1}) becomes 
\beqn
\frac{\partial}{\partial t} \, \Delta_{_T} \widetilde q^{\, \, \pm} (x,t) =
\frac{\alpha_s (t)}{2\pi} \,  
\Delta_{_T} \widetilde P_{q^\pm} (x) \otimes \Delta_{_T} 
\widetilde q^{\, \, \pm} (x,t)
\ .
\label{eqn:DGLAP2}
\eeqn

\section{{\bf Simpson method}}\label{SIMPSON}
\setcounter{equation}{0}
\setcounter{figure}{0}
\setcounter{table}{0}

There are various numerical methods for solving the DGLAP
equations \cite{BGR}.
We have investigated two methods, the Laguerre-polynomial \cite{LAG}
and the brute-force \cite{BF} methods.
The splitting functions and parton distributions are expanded by
the Laguerre polynomials in the first method.
Even though the Laguerre method has an advantage of computing time,
it may not be very accurate 
in the spin-independent nonsinglet case at small $x$.
In the brute-force method, the variables $t$ and $x$ are divided 
into small steps, then the integral is calculated in the simplest manner.
This method is accurate if a large number of steps is taken.
However, it inevitably takes much computing time.
In order to resolve these problems, the Simpson method \cite{SIMP}
is employed in the $z$ integration of Eq. (\ref{eqn:conv}).
Here, we report its numerical analysis.

In the following numerical solution, the brute-force method
is still used in the variable $t$ and the Simpson method
is introduced in calculating the $z$ integration.
The brute-force method for solving a differential equation is 
often called Euler method \cite{SIMP},
so that we use this terminology in this paper.
The variables $t$ and $z$ are divided into $N_t$ and $N_z$ steps,
then differentiation and integration are defined by
\begin{align}
\frac{d f(t)}{dt} =& \, \frac{f(t_{j+1})-f(t_j)}{\delta t}
\ ,
\label{eqn:EULER} \\
\int g(z) \ dz =& \sum_{k=2}^{N_z} \, \frac{\delta z}{3} \, 
[ \, g(z_{k-1}) + 4 g(z_k) + g(z_{k+1}) \, ] \ ,
\nonumber \\ &  where \ \ k = 2, \, 4, \, \cdot \cdot \cdot, 
                \, N_z \ \ \ \ \ .
\label{eqn:SIMPSON}
\end{align}
The $k$ summation is taken over the even numbers from 2 to $N_z$,
so that $N_z$ has to be an even number.

With these replacements, the evolution equation can be solved
rather easily. Equation (\ref{eqn:DGLAP2}) is written in the following form:
\begin{align}
\Delta_{_T} \widetilde q^{\, \, \pm} (x_m,t_{j+1}) =& \, 
  \Delta_{_T} \widetilde q^{\, \, \pm} (x_m,t_j) + 
  \delta t \, \frac{\alpha_s (t_j)}{2\pi} \, 
  \sum_{k=2m}^{N_z} \frac{\delta z}{3 z_k} \, F(z_k, x_m, t_j)
\nonumber \\
& k = 2m, \, 2m+2, \, \cdot \cdot \cdot, \, N_z \ \ \ \ \
\ ,
\end{align}
where the variable $x$ is divided into $N_x = N_z / 2$ steps.
The function $F$ is defined by the following:
\begin{align}
F(z_k, x_m, t_j) = \, & 
\Delta_{_T} \widetilde P_{q^\pm} (z_{k-1}) \, 
\Delta_{_T} \widetilde q^{\, \, \pm} 
\left( \frac{x_m}{z_{k-1}}, t_j \right)
+ 4 \ \Delta_{_T} \widetilde P_{q^\pm} (z_k) \, 
\Delta_{_T} \widetilde q^{\, \, \pm} \left( \frac{x_m}{z_k}, t_j \right)
\nonumber \\
& \ + \Delta_{_T} \widetilde P_{q^\pm} (z_{k+1}) \, 
\Delta_{_T} \widetilde q^{\, \, \pm} \left( \frac{x_m}{z_{k+1}}, t_j \right)
\ .
\label{eqn:SIMPSON2}
\end{align}
First, the evolution from $t_1 = \ln Q_0^2$ to $t_2 = t_1 + \delta t$ is
calculated in the above equation by providing the initial distribution 
$\Delta_{_T} \widetilde q^{\, \, \pm} (x_m/z_k, t_{j=1})$.
Repeating this step $N_t-1$ times, we obtain the final distribution
at $t = t_{N_t+1}$.
However, the small $x$ region is often important practically.
If the accurate results are required in the small $x$ region,
$x$ and $z$ should be replaced by $\log_{10} x$ and $\log_{10} z$.
Then, the evolution equation Eq. (\ref{eqn:DGLAP2}) is replaced by
\beqn
\frac{\partial}{\partial t} 
\Delta_{_T} \widetilde q^{\, \, \pm} (\log_{10} x,t) = 
\ln 10 \ \frac{\alpha_s (t)}{2\pi} 
\int_{\log_{10} x}^{0} \! \! \! \!  \! \! \! \! 
d(\log_{10} z) 
\Delta_{_T} \widetilde P_{q^\pm} (\log_{10} z) \Delta_{_T} 
\widetilde q^{\, \, \pm} \left(\log_{10} \frac{x}{z}, t \right) 
\, .
\label{eqn:DGLAP3}
\eeqn
In order to apply the Simpson method to this equation,
$\log_{10} x$ and $\log_{10} z$ are divided into $N_x$ and $N_z$ steps.
Our FORTRAN code supports both the linear-step and the
logarithmic-step cases,
and the details are explained in the following sections.

\vfill\eject
\section{{\bf Description of input parameters and initial \\ distribution}}
\label{INPUT}
\setcounter{equation}{0}
\setcounter{figure}{0}
\setcounter{table}{0}

For running the FORTRAN-77 program H1EVOL, a user should supply
seventeen input parameters from the file \#10.
In addition, an input distribution(s) should be
given in a function subroutine(s) in the end of the FORTRAN program
or in an input data file(s).
The $i$-th initial distribution is written 
in the output file \#30+$i$ (\#31 -- \#38). 
Evolution results for the distribution-$i$ are written
in the output file \#20+$i$ (\#21 -- \#28). 
The maximum number of the input distributions is eight.
We explain the input parameters and the initial distributions
in the following.

\subsection{Input parameters}\label{PARAMET}

There are following seventeen parameters.

\vspace{0.3cm}

\small
\noindent
\begin{tabular}{|l|c|l|} \hline
IREP    & 1 & finish the program after this evolution
\\ \cline{2-3}
        & 2 & proceed to the next evolution with new input parameters
\\ \hline \hline
IOUT   & 1 & write $x$ and $x\Delta_{_T} q(x)$ at fixed $Q^2$ (=Q2) 
             in the file(s) \#21 -- \#28
\\ \cline{2-3}
       & 2 & write $Q^2$ and $x\Delta_{_T} q(Q^2)$ at fixed $x$ (=XX) 
             in the file(s) \#21 -- \#28
\\ \hline \hline
IREAD  & 1 & give initial distribution(s) in function subroutine(s)
\\ \cline{2-3}
       & 2 & give initial distribution(s)
             from the file(s) \#11 -- \#18
\\ \hline \hline
INDIST & 1 & do not write initial distribution(s)
\\ \cline{2-3}
       & 2 & write initial distribution(s) in the file(s) \#31 -- \#38
\\ \cline{2-3}
       & 3 & write initial distribution(s) in the file(s) \#31 -- \#38 \\
       &   & without calculating evolution
\\ \hline \hline
IORDER   & 1 & leading order (LO) in $\alpha_ s$
\\ \cline{2-3}
         & 2 & next-to-leading order (NLO)
\\ \hline\hline
IMORP    & 1 & $x\Delta_{_T} q_i - x\Delta_{_T} \bar q_i$ type distribution
\\ \cline{2-3}
         & 2 & $x\Delta_{_T} q_i + x\Delta_{_T} \bar q_i$ type distribution
\\ \hline\hline
ILOG    & 1 & linear-$x$ and linear-$z$ steps are taken 
              in calculating the evolution \\
        &   & and in writing the $x$-dependent output. 
\\ \cline{2-3}
        & 2 & logarithmic-$x$ and logarithmic-$z$ steps \\
        &   & (We recommend to use ILOG=2 in a general case.)
\\ \hline
\end{tabular}


\noindent
\begin{tabular}{|l|l|} \hline
Q02   &  initial $Q^2$ ($\equiv Q_0^2$ in GeV$^2$ )
                   at which an initial distribution is supplied
\\ \hline
Q2    &  $Q^2$ to which the distribution is evolved 
         ($Q^2\ne Q_0^2$, $Q^2 > Q_0^2$ or $Q^2 < Q_0^2$)
\\ \hline
DLAM  &  QCD scale parameter $\Lambda_{QCD}$ in GeV
\\ \hline
NF    &   number of quark flavors $N_f$
\\ \hline
XX    &  $x$ at which $Q^{2}$ dependent distributions are written
(IOUT=2 case)
\\ \hline
NX    &  number of $x$ steps $N_x$ (NX $\le$ 3000)
\\ \hline
NT    &  number of $t$ steps $N_t$ (NT $\le$ 3000)
\\ \hline
NSTEP &  number of $x$ steps or $t$ steps for writing output distribution(s) 
\\ \hline
XMIN & minimum of $x$, $x_{min}$ (0 $<$ XMIN $<$ XX)
\\ \hline
NFI & number of distributions which are evolved simultaneously
(NFI $\le$ 8)
\\ \hline
\end{tabular}
\normalsize

\vspace{0.5cm}
\noindent
Numerical values of the parameters should be supplied
in the file \#10, then these are read in the program
through the subroutine GETPA1.
{\it It should be noted that the first four lines of 
     the file \#10 are comments.}
Therefore, the parameters are given
from the fifth line as shown in the following example.
There are irrelevant parameters in some cases:
\beqn
IOUT=1 & case & \cdot \cdot \cdot \ \ XX
\nonumber \\
INDIST=3 & case & \cdot \cdot \cdot \ \ 
                 IORDER, IMORP, Q2, DLAM, NF, XX, NT
\nonumber \\
IORDER=1 & case & \cdot \cdot \cdot \ \ IMORP \ \ .
\nonumber
\eeqn
Although they are not used in running the evolution part,
their numerical values should be supplied in the file \#10
within the allowed ranges.

The meaning of IREAD is explained in Section 4.2.
The parameter IMORP indicates a plus or minus type distribution
$\sum_i a_i x (\Delta_{_T} q_i \pm \Delta_{_T} \bar q_i)$,
where $a_i$ are some constants.
In the NFI $\ge$ 2 case, distributions with
the same $\pm$ type can be evolved at the same time.
The maximum number of the distributions is set up as eight in the program. 
For example, if one would like to evolve two $\Delta_{_T}q^+$
type initial distributions at $Q^2$=4.0 GeV$^2$ to the distributions
at $Q^2$=200.0 GeV$^2$ by the NLO DGLAP equation with $N_f$=4 
and $\Lambda$=0.231 GeV, the input parameters could be
IREP=1, IOUT=1, IREAD=1, INDIST=1, IORDER=2, IMORP=2, ILOG=2,
Q02=4.0, Q2=200.0, DLAM=0.231, NF=4, XX=0.1,
NX=500, NT=50, NSTEP=50, XMIN=0.0001, and NFI=2.
In this case, the input file \#10 is:

\vspace{0.3cm}
\noindent
\ \ \ \ \ \ ---------------------------------------------------------------------

\noindent
\ \ \ \ \ \  IREP,IOUT,IREAD,INDIST,IORDER,IMORP,ILOG

\noindent
\ \ \ \ \ \  Q02,Q2,DLAM,NF,XX,NX,NT,NSTEP,XMIN,NFI

\noindent
\ \ \ \ \ \ ---------------------------------------------------------------------

\noindent
\ \ \ \ \ \ 1, 1, 1, 1, 2, 2, 2 

\noindent
\ \ \ \ \ \ 4.0, 200.0, 0.231, 4, 0.1, 500, 50, 50, 0.0001, 2

\vspace{0.3cm}
\noindent
The first four lines are comments.
The parameter values from IREP to ILOG should be written in the fifth
line and the remaining ones in the next line.
If one would like to repeat the evolution, for example 
if the LO evolution is also needed, the input file is the following:

\vspace{0.3cm}
\noindent
\ \ \ \ \ \ ---------------------------------------------------------------------

\noindent
\ \ \ \ \ \  IREP,IOUT,IREAD,INDIST,IORDER,IMORP,ILOG

\noindent
\ \ \ \ \ \  Q02,Q2,DLAM,NF,XX,NX,NT,NSTEP,XMIN,NFI

\noindent
\ \ \ \ \ \ ---------------------------------------------------------------------

\noindent
\ \ \ \ \ \ 2, 1, 1, 1, 2, 2, 2 

\noindent
\ \ \ \ \ \ 4.0, 200.0, 0.231, 4, 0.1, 500, 50, 50, 0.0001, 2

\noindent
\ \ \ \ \ \ 1, 1, 1, 1, 1, 2, 2 

\noindent
\ \ \ \ \ \ 4.0, 200.0, 0.231, 4, 0.1, 500, 50, 50, 0.0001, 2

\vspace{0.3cm}
\noindent
In the same way, one can repeat the evolution further by choosing IREP=2.
The evolved results are written continuously in the same output file(s).

\subsection{Initial distributions supplied by function subroutines \\
(IREAD=1)}
\label{INPUTFUN1}

If IREAD=1 is chosen, an input distribution(s) at $Q_0^2$
should be supplied in the end of the FORTRAN program H1EVOL
as a function subroutine(s) H1IN1(X) -- H1IN8(X).
For example, three distributions 
(e.g. $x \Delta_{_T} u^+$, $x \Delta_{_T} d^+$, and $x \Delta_{_T} s^+$)
can be evolved at the same time in the NFI=3 case.
These initial distributions should be supplied in the 
function subroutines H1IN1(X), H1IN2(X), and H1IN3(X).
The remaining functions H1IN4(X) -- H1IN8(X) may be set to zero. 
The evolution results for the initial distribution-i, which is given
by the function H1INi(X) (i=1, 2, $\cdot \cdot \cdot$, NFI), are written 
in the output file \#20+i.
If INDIST=2 or 3 is chosen, the initial distribution(s) 
is written in the file \#30+i.

To be exact, there is no information on the input transversity
distribution because experimental data do not exist. 
We discuss this point in Section 6.1. In our program, 
the Gehrmann-Stirling (GS) type-A \cite{GS} singlet distribution 
$x \Delta_{_T} q_{_S} \equiv 
\sum_i^{N_f} (x \Delta_{_T} q_i + x \Delta_{_T} \bar q_i)$ 
is given in the function H1IN1(X), and the distribution
$[(x \Delta_{_T} u + x  \Delta_{_T} \bar u) 
 - (x \Delta_{_T} d + x  \Delta_{_T} \bar d)]/2$ 
is given in H1IN2(X).
Additionally, the valence-quark distribution
$x \Delta_{_T} u_{_V} + x \Delta_{_T} d_{_V}$ is given 
in the function H1IN1(X) and the distribution
$[(x \Delta_{_T} u - x  \Delta_{_T} \bar u) 
 - (x \Delta_{_T} d - x  \Delta_{_T} \bar d)]/2$ 
is given in H1IN2(X) as comments.

\subsection{Initial distributions supplied by data files \\ (IREAD=2)}
\label{INPUTFUN2}

If IREAD=2 is chosen, an input distribution(s) at $Q_0^2$
should be supplied in a separate data file(s) \# 11 -- \#18
as shown in the following.
We give the singlet distribution as an example.

\begin{tabbing}
      0.000100 \ \ \ \ \ \ \= 0.001721 \kill
      0.000100   \>  0.001721 \\
      0.000110   \>  0.001612 \\
      0.000120   \>  0.001485 \\
      0.000132   \>  0.001339 \\
      0.000145   \>  0.001172 \\
       ...       \>   ...     \\
       ...       \>   ...     \\
       ...       \>   ...     \\
      1.000000   \>  0.000000 \\
\end{tabbing}

\vspace{-0.3cm}
The first column is the $x$ values and the second
one is the corresponding distribution values.
The data should be sorted from the smallest $x$ to larger ones, 
and the data at $x \le x_{min}$ and at $x$=1.0 must be supplied.
The last line number should be smaller than three thousand.
The evolution results for the initial distribution
in the file \#10+i (i=1, 2, $\cdot\cdot\cdot$, NFI)
are written in the file \#20+i.
If INDIST=2 or 3 is chosen, the initial distribution(s)
at NSTEP+1 points is written in the file(s) \#30+i.

\section{{\bf Description of the program H1EVOL}}\label{H1EVOL}
\setcounter{equation}{0}
\setcounter{figure}{0}
\setcounter{table}{0}

\subsection{Main program H1EVOL}
\label{MAIN}
 
The seventeen input parameters are read by calling
the subroutine GETPA1 and these parameters are checked 
by the subroutine ERR1.
If there is an error, the program writes the error message
in the output file \#6 and stops.
Color constants and other necessary constants are calculated
in the subroutine GETPA2.
Then, $t$, $x$, $z$, and the splitting function are evaluated
at each step point by the subroutines GETT, GETX, GETZ, and GETP
respectively.
If IOUT=2 is chosen, the XMIN is replaced by XX and $x$ values 
in the array X(I) are shifted in the subroutine GETMIN so as
to have X(1)=XX, $\cdot\cdot\cdot$, X(NX+1)=1.0.
The initial distribution(s) is calculated in the subroutine GETINI
or is read from the data file(s) by the subroutine REDINI,
then it is stored in the array H1(IN,I).
Here, the argument IN is the distribution number from 1 to NFI.
The initial distribution(s) is written in the file(s) \#30+IN
by the subroutine OUTXD if INDIST$\ne$1.

Since all the necessary values are ready in the above procedures,
we start the evolution calculation.
First, the initial distribution(s) is stored in the array WH1(IN,I) 
and $\alpha_s$ at $t_1$ is calculated by the subroutine GETALP.
The evolved distribution(s) at $t_2 = t_1 + \delta t$ is calculated 
by the subroutines GETINT and GETH1, and the results are stored
in H1(IN,I). 
If IOUT=2, the result(s) at $x$ = XX is stored in the array H1Q2(IN,I).
Next, the evolution results are stored in WH1(IN,I) and $\alpha_s$ at
$t_2$ is evaluated. Then, the evolution 
from $t_2$ to $t_3 = t_2 + \delta t$ is calculated in the same way.
Repeating this step,
we obtain the final distribution(s) at $t = t_{N_t+1}$.
The calculated values are interpolated either in $t$ or in $x$
depending on IOUT by using the subroutine SPLINE and the function SEVAL.
The final results at NSTEP+1 points are written in the output 
file(s) \#10+IN by the subroutines OUTXD or OUTQ2D.

\subsection{Subroutines GETPA1(IREP,IEORD,IERR), GETPA2, \\
and ERR1(IREP,IERR)}
\label{GETPA}

The subroutine GETPA1 reads the seventeen input parameters 
from the input file \#10. These parameters are checked
whether they are within the valid ranges by the subroutine ERR1. 
If there is an error, IERR=1 is returned from the subroutine ERR1
and the program stops with warning notification in the file \#6.
The subroutine GETPA2 calculates necessary color constants, 
$\pi$, $\beta_0$, and $\beta_1$.

\subsection{Subroutines GETT(Q02,Q2,NT,T,DT) \\
and GETX(XMIN,XMAX,NX,ILOG,X,DX)}
\label{GETTX}

These subroutines evaluate the values of $t = \ln Q^2$ and $x$
at $N_t+1$ and $N_x+1$ points, and they are 
stored in the arrays T(I) with I=1, 2, $\cdot\cdot\cdot$, $N_t+1$
                 and X(I) with I=1, 2, $\cdot\cdot\cdot$, $N_x+1$.
In the subroutine GETX, either the linear-$x$ (ILOG=1 case) 
or logarithmic-$x$ (ILOG=2 case) step is chosen.
These subroutines also return the value 
of $\delta t = [t(Q^2)-t(Q_0^2)]/N_t=DT$ and $\delta x=DX$,
where $\delta x$ is defined by $\delta x = (1-x_{min})/N_x$
in the ILOG=1 case
or $\delta x = (0 - \log_{10} x_{min})/N_x$ in the ILOG=2 case. 

\subsection{Subroutine GETMIN(XX,X,NX,XMIN)}
\label{GETMIN}

In the IOUT=2 case, the subroutine GETMIN is called after getting X(I).
This subroutine replaces the XMIN value by XX because
the distribution information at $x<XX$ is not necessary
in the evolution calculation.
Then, the values in the array X(I) are shifted.
For example, if XX is in the region X(M) $<$ XX $<$ X(M+1), 
X(1) is replaced by XX 
and X(2), X(3), $\cdot \cdot \cdot$, X(NX$-$M+2) are replaced by
X(M+1), X(M+2), $\cdot \cdot \cdot$, X(NX+1). 
The number of steps $N_x$ is also replaced by $N_x-M+1$.

\subsection{Subroutine GETZ(ILOG,X,NX,DX,Z,NZ,DZ)}
\label{GETZ}

The subroutine GETZ calculates the values of $z$ at $N_z+1$ 
$(N_z= 2 N_x)$ points
and stores these values into the array Z(K), K=1, 2, $\cdot\cdot\cdot$, 
$N_z+1$. The first one is given by $\rm Z(1)=X(1)$, and the second one
Z(2) is calculated by
${\rm Z(2)} = {\rm Z(1)} + [{\rm X(2)} - {\rm X(1)}]/2$ in ILOG=1 case or
${\rm Z(2)} = {\rm Z(1)} + [\log_{10}{\rm X(2)} - \log_{10}{\rm X(1)}]/2$ 
in ILOG=2 case.
The subsequent ones are given by 
Z(3)=X(2), Z(4)=Z(3)+$\delta x$/2, Z(5)=X(3), 
           Z(6)=Z(5)+$\delta x$/2, $\cdot\cdot\cdot$ in the ILOG=1 case
or by
Z(3)=X(2), $\log_{10}$Z(4)=$\log_{10}$Z(3)+$\delta x$/2, Z(5)=X(3), 
           $\log_{10}$Z(6)=$\log_{10}$Z(5)+$\delta x$/2, 
$\cdot\cdot\cdot$ in the ILOG=2 case.

\subsection{Subroutine GETP(IORDER,SIGN)}
\label{GETP}

The subroutine GETP calculates the LO and NLO splitting functions
at the $N_z+1$ points and stores the calculated values
in the arrays P0(K), P0PF(K), P1(K) and P1PF(K).
The splitting functions are calculated by the functions PLO, P0PL, 
PNLO, and P1PL.
Additionally, the extra factors associated with the `+' function
or $\delta$ function are also evaluated by calling the subroutines
P0DL and P1DL.

\subsection{Subroutines GETINI(NFI,H1) and REDINI(NFI,H1,IERR)}
\label{GETINI}

If IREAD=1 is chosen, the subroutine GETINI calculates the initial 
distribution(s) by calling the function(s) H1IN1(X) -- H1IN8(X). 
The subroutine REDINI reads the initial distribution(s) from the 
input data file(s) and interpolates the distributions(s) 
by using SPLINE and SEVAL if IREAD=2.

\subsection{Subroutine GETALP(T,IORDER,ALP2PI)}
\label{GETALP}

The subroutine GETALP calculates the $\alpha_s / 2\pi$=ALP2PI at given $t$.

\subsection{Subroutines GETINT(DINTEG), FUN(K,I,IN), \\
and SIMPS(FUN,I,IN,DZ)}
\label{GETINTEG}

The subroutine GETINT calculates the integral by calling the subroutine
SIMPS. The SIMPS integrates the integrand FUN
by using the Simpson method. 
The integral results in the array DINTEG are then returned.
The factor IN is the input distribution number, IN=1, 2, $\cdot\cdot\cdot$,
or NFI. 
The factors K and I represent the $z$ and $x$ positions, Z(K) and X(I).

\subsection{Subroutine GETH1(DT1,DINTEG,H11)}
\label{GETH1}

The subroutine GETH1 calculates the distribution at $t=t_j$ from the
one at $t=t_{j-1}$ with the integral obtained
by GETINT. The evolution results are stored in H11(IN,I).
The factor DT1 is the $t$ step.

\subsection{Subroutines OUTXD(ILOG,H1,NFILE) \\
and OUTQ2D(Q2T,H1Q2)}
\label{OUTPUT}

These subroutines interpolate the final evolution results 
and write the distribution(s) at NSTEP+1 points in the output file(s).
The OUTXD (OUTQ2D) writes the $x$ ($Q^2$) dependent results.
The OUTXD is also used for writing the initial distribution(s)
in the INDIST=2 or 3 case. The NFILE is the factor for writing
either the initial distribution (NFILE=1)
or the final one (NFILE=2).

\subsection{Subroutine SPLINE(N,X,Y,B,C,D,IN) \\
and function SEVAL(N,XX,X,Y,B,C,D,IN)}
\label{SPLINE}

The subroutine SPLINE calculates the coefficients, B(IN,I), C(IN,I),
and D(IN,I) ($I=1,2, \cdot\cdot\cdot, N$) by the cubic Spline interpolation
for the distribution type IN. The interpolated points Y(IN,I) at X(I) 
($I=1,2, \cdot\cdot\cdot, N$) should be supplied.
This interpolation program is taken from Ref. \cite{FMM}.
Using the obtained Spline coefficients, the function SEVAL calculates 
the value of Y at given XX. 

\subsection{Functions PLO(Z), P0PL(Z), PNLO(Z,SIGN), \\
and P1PL(Z)}
\label{SPLIT1}

The functions PLO and PNLO calculate the LO and NLO splitting functions. 
The functions P0PL and P1PL calculate the LO and NLO 
`+' function parts in the splitting functions. If IMORP=1 (SIGN=$-$1.0), 
the PNLO calculates the $\Delta_{_T} q - \Delta_{_T} \bar{q}$ type
splitting function, and it calculates the 
$\Delta_{_T} q + \Delta_{_T} \bar{q}$ type if IMORP=2 (SIGN=+1.0). 

\subsection{Subroutines P0DL(PLUS0,DEL0) and P1DL(PLUS1,DEL1) }
\label{SPLIT2}

The subroutines P0DL and P1DL calculate the extra factors which
are related to the `+' function [see the second term
of Eq. (\ref{pfun2})] and the $\delta$ function.
These factors are taken out from the integral.

\subsection{Functions S2(Z), SPENCE(X), and DGAUSS(EXTERN,A,B)}
\label{S2}

The function S2 calculates $S_2(x)$ which appears in
the splitting function.
The $S_2(x)$ is expressed in terms of the Spence function
in Eq. (\ref{eqn:S2}). The Spence function is defined by the integral form 
in Eq. (\ref{eqn:SPENCE}), 
and it is numerically calculated by the function DGAUSS 
in the Gauss-Legendre method with the integrand SPENCE.

\subsection{Function GS(X,A,B,C,D,E,F)}
\label{GS}

This function calculates the GS type parton distribution
with the parameters A, B, C, D, E, and F as
$A B x^C (1 - x)^D (1 + E x +F \sqrt x)$.

\subsection{Functions H1IN1(X), H1IN2(X), $\cdot \cdot \cdot$, and H1IN8(X)}
\label{H1IN}

These functions calculate the initial distributions.
For example, three initial distributions should be
provided in the NFI=3 case by H1IN1(X), H1IN2(X), and H1IN3(X),
and the remaining functions should be set to zero.

\vfill\eject
\section{{\bf Numerical analysis}}\label{ANALYSIS}
\setcounter{equation}{0}
\setcounter{figure}{0}
\setcounter{table}{0}

\subsection{Accuracy of $\bf Q^2$ evolution results}
\label{ACCURA}

There are two parameters which determine the accuracy of our numerical
results. They are the numbers of steps: $N_x$ and $N_t$.
Another number of steps $N_z$ is fixed by the relation $N_z = 2 N_x$.
In this subsection, we discuss how the evolution results
depend on these parameters
and how long it takes for calculating the evolution.

There is a practical problem in the transversity evolution
in the sense that there is no input distribution.
For example, the input parton distributions are obtained by
fitting various experimental data in the unpolarized
distribution studies. At this stage, there is no experimental
information on the transversity distribution, so that 
the initial distribution could not be supplied. 
An item of good news is that the lattice QCD studies started
producing a $x$ distribution \cite{GHPRS3}
although it is still somewhat away from the very reliable one. Therefore,
we may use the initial distribution calculated by a bag model \cite{H1},
a nonrelativistic quark model \cite{SV,MSS}, or 
the Gl\"uck-Reya-Stratmann-Vogelsang
distribution \cite{GRSV,BCD} at very small $Q^2$. 
It is known that the transversity distribution and
the corresponding longitudinally polarized one are almost 
the same at small $Q^2$ according to Ref. \cite{H1}. 
However, the applicability of perturbative QCD becomes dubious
in the small $Q^2$ region.
Therefore, instead of stepping into the nonperturbative region,
we decided to use a longitudinally polarized distribution
at larger $Q^2$ as the input transversity distribution.
The Gehrmann-Stirling (GS) type-A distributions \cite{GS} are used
throughout this paper. They are given at $Q^2$=4 GeV$^2$. 
We warn the reader that the assumption
$\Delta q=\Delta_{_T} q$ would not be valid in the $Q^2$=4 GeV$^2$ 
region even though it would work at small $Q^2$ ($\sim \Lambda_{QCD}^2$).
Future experimental data will clarify the difference between
the two distributions.

We show the dependence on $N_x$ and $N_t$ by taking
the singlet distribution.
First, we fix the $x$-step number at $N_x$=1000, then the $t$-step 
number $N_t$ is varied as 10, 50, 200, and 1000. 
The input distribution is the GS-A singlet distribution
at $Q^2$=4 GeV$^2$ \cite{GS}. 
Because there is little information on each antiquark distribution \cite{SK},
flavor symmetric distributions are assumed for the antiquark distributions.
Calculated evolution results for the distribution
$x \Delta_{_T} q_{_S} \equiv x \sum_i^{N_f} ( \Delta_{_T} q_i 
                                     + \Delta_{_T} \bar q_i) $
are shown in Fig. \ref{fig:nt}.
We use $N_f=$4 and $\Lambda$=231 MeV in calculating the evolution.
The GS-A singlet distribution is evolved to the one at $Q^2$=200 GeV$^2$ 
by our NLO program.
The initial distribution is supplied at the end of
the program as the subroutine H1IN1(X).
The input parameters are 
IREP=1, IOUT=1, IREAD=1, INDIST=1, IORDER=2, IMORP=2, ILOG=2,
Q02=4.0, Q2=200.0, DLAM=0.231, NF=4, XX=0.1,
NX=1000, NT=10, 50, 200, or 1000, NSTEP=100, XMIN=0.0001, and NFI=1.
There is almost no difference between the various results 
in Fig. \ref{fig:nt} except for the $N_t$=10 case.
It means that merely fifty steps are enough for the getting
the accurate evolution. The small number of steps is enough
because the scaling violation is a small logarithmic effect. 

Next, the step $N_x$ is varied with fixed $N_t$=100 in Fig. \ref{fig:nx},
where $N_x$=50, 200, 500, and 3000 are taken. 
Numerical accuracy is slightly worse in the $N_x$=50 case; however,
the accurate results are obtained by taking the five hundred steps.
Defining the evolution accuracy by 
$| \, [\Delta_{_T} q (N_x,N_t) - \Delta_{_T} q (N_x=3000,N_t=1000) ]
 / \Delta_{_T} q (N_x=3000,N_t=1000) \, |$,
we find that the accuracy is better than 1\% 
in the region $10^{-5}<x<0.8$ with $N_x=500$ and $N_t=50$
in the logarithmic-$x$ steps (ILOG=2).
If one is more interested in the large-$x$ region,
one had better take the linear-$x$ steps (ILOG=1).
However, we recommend the user to use the logarithmic steps
in a general case.

Finally, we discuss a typical running CPU time.
If one initial distribution is evolved with $N_x=500$ and $N_t=50$, 
it is about forty seconds on the AlphaServer 2100 4/200
and two and a half minutes on the SUN-IPX.
From these analyses, we find that our program can be run 
on a typical workstation or possibly on a personal computer
without spending much computing time.
However, if one would like to calculate the evolution repeatedly 
or to evolve several distributions simultaneously, 
a reasonably powerful machine is recommended.

\subsection{$\bf Q^2$ evolution of transversity distributions}
\label{RESULTS}

There are some studies on the transversity evolution \cite{BCD,SV,BST,MSSV}; 
however, it is not a well investigated topic. Therefore, we show evolution
results of various transversity distributions in this subsection.
In particular, the LO and NLO evolution differences are shown, 
then they are compared with those of 
the corresponding longitudinally polarized distributions
by using the BFP1 program \cite{BF}.
The transversity NLO evolution is the same in the $MS$ \cite{KM,HKK}
and $\overline{MS}$ \cite{V} schemes. 

First, the singlet evolution results are shown in Fig. \ref{fig:sing}.
The same GS-A distribution is assumed for the transversity and 
longitudinally-polarized parton distributions at $Q^2$=4 GeV$^2$.
Furthermore, the LO and NLO distributions are assumed the same in
both cases. The initial distribution is shown by the dotted curve.
It is evolved to the distributions at $Q^2$=200 GeV$^2$ by the 
transverse or longitudinal evolution equation.
The transversity NLO effects increase the evolved distribution
at medium-large $x$ and also at small $x$ ($<$0.01), 
and they decrease the distribution in the intermediate
$x$ region ($0.01<x<0.1$). The NLO contributions are rather
different from those of the longitudinal evolution.
The transversity evolution differs significantly from
the longitudinal one either in the LO case or in the NLO.
The evolved transversity distribution $\Delta_{_T} q_s$
is significantly smaller than the longitudinal one $\Delta q_s$
in the region $x\sim 0.1$. The magnitude of $\Delta_{_T} q_s$
itself is also smaller than that of $\Delta q_s$ at very small $x$
($<0.07$). Therefore, one should be careful about the $Q^2$
evolution difference between the two distributions in extracting
information on $\Delta_{_T} q_s$ from experimental data.
The explicit $Q^2$ dependence is shown in Fig. \ref{fig:q2},
where the distributions are shown at $x$=0.001, 0.01, 0.1, and 0.5.

In parametrizing the parton distributions \cite{RHIC-SPIN-J}, 
it is natural to provide the valence and sea quark distributions
separately. We expect that future parametrizations on $\Delta_{_T} q_s$
will be given in the same manner. 
Therefore, it is important to show the evolution results of 
a valence-quark distribution.
The evolution calculations are done in the same way as those
in Fig. \ref{fig:sing} except that the initial distribution
is the nonsinglet one $x (\Delta u_v+\Delta d_v)=
                       x (\Delta_{_T} u_v+\Delta_{_T} d_v)$.
The evolution results are shown in Fig. \ref{fig:val}.
The dotted curve is the initial distribution, and evolved
distributions are solid and dashed curves.
The evolved transversity distributions are significantly
smaller than the longitudinally polarized ones, in particular
at small $x$.

The other interesting distribution is the flavor asymmetric
distribution $\Delta_{_T} \bar u - \Delta_{_T} \bar d$.
Although the unpolarized version $\bar u - \bar d$ is rather
well known \cite{SK} due to the measurements of
the Gottfried-sum-rule violation and
the Drell-Yan p-n asymmetry, the longitudinally polarized
one $\Delta \bar u - \Delta \bar d$ is not investigated experimentally.
Because of the difference between the splitting functions
for the $\Delta_{_T} q+ \Delta_{_T}\bar q$ and 
$\Delta_{_T} q- \Delta_{_T}\bar q$ type distributions,
a finite $\Delta_{_T} \bar u - \Delta_{_T} \bar d$ could be obtained
through the $Q^2$ evolution \cite{MSSV} as it appears
also in the unpolarized case \cite{SK}.
We show the evolution results of $\Delta_{_T} \bar u - \Delta_{_T} \bar d$
and $\Delta \bar u - \Delta \bar d$ in Fig. \ref{fig:ubdb}.
As it is obvious from the figure, the obtained distributions
are fairly small. In order to get the accurate evolution in 
Fig. \ref{fig:ubdb}, large numbers of $x$ and $t$ steps should be
chosen: for example $N_x$=3000 and $N_t$=500.
There is no flavor asymmetry in the GS-A antiquark distributions,
so that the initial distributions are
$\Delta_{_T} \bar u - \Delta_{_T} \bar d =
      \Delta \bar u - \Delta      \bar d =0$.
Because the difference between the $q^\pm$ splitting functions
does not appear in the LO, the evolved distributions show no asymmetry
$(\Delta_{_T} \bar u - \Delta_{_T} \bar d)_{LO} =
      (\Delta \bar u - \Delta      \bar d)_{LO} =0$
at $Q^2$=200 GeV$^2$.
The situation is completely different in the NLO.
The NLO evolution results are shown by the solid and dashed curves.
It is interesting to find the finite distributions
in the NLO evolution. The evolved transversity distribution
$x(\Delta_{_T} \bar u - \Delta_{_T} \bar d)$ is concentrated
in the $x\sim 0.04$ region, and the longitudinal one
is distributed in the smaller $x$ ($\sim$0.004) region.
However, we should note the magnitude. The distributions
are typically 0.0001, which is too small to be found
experimentally in the near future. 
The situation is the same in the unpolarized: 
the perturbative QCD correction to the Gottfried sum is 
typically 0.3\% with respect to the 30\% experimental violation.
The $\Delta_{_T} \bar u - \Delta_{_T} \bar d$ as well as
    $\Delta \bar u - \Delta      \bar d$ will be measured
in the W-charge-asymmetry experiments at RHIC \cite{WASYM}.
Even if finite distributions are found, they are unlikely
to be explained by the perturbative mechanism.

\vfill\eject
\section{{\bf Summary}}\label{SUMMARY}
\setcounter{equation}{0}
\setcounter{figure}{0}
\setcounter{table}{0}

We investigated numerical solution of the $Q^2$ evolution
equation for the transversity distribution $\Delta_{_T} q$ 
or the structure function $h_1$.
A useful FORTRAN program was created for calculating the
evolution in the LO and NLO cases. The renormalization
scheme is $MS$ or $\overline{MS}$ in the NLO.
The numerical solution was obtained by dividing the variables
$x$ and $t=\ln\, Q^2$ into small steps with the numbers of steps,
$N_x$ and $N_t$. The Euler and Simpson methods are used for solving
the integrodifferential equation.
We find that the evolution accuracy is better
than 1\% with $N_x \ge 500$ and $N_t \ge 50$
in the $x$ region $10^{-5}<x<0.8$\, .

Using the program, we showed the evolution results of 
the valence-quark distribution $x(\Delta_{_T} u_v + \Delta_{_T} d_v)$,
the singlet distribution 
$x\sum_i^{N_f} ( \Delta_{_T} q_i + \Delta_{_T} \bar q_i )$,
and the flavor asymmetric distribution 
$x(\Delta_{_T} \bar u - \Delta_{_T} \bar d)$.
The evolution results are compared with those
calculated by the longitudinal DGLAP evolution equations.
Because the transverse and longitudinal evolution results
are significantly different, we should be careful in
analyzing experimental data.
The $Q^2$ variations of the flavor asymmetric distribution
are fairly small so that a nonperturbative mechanism should be
explored if a finite distribution is found experimentally.
The transversity evolution of these distributions
should be tested by future experimental projects such
as the RHIC-Spin.

\vspace{0.7cm}
\section*{{\bf Acknowledgment}}
\addcontentsline{toc}{section}{\protect\numberline{\S}{Acknowledgments}}

MH, SK, and MM thank the Research Center for Nuclear Physics 
in Osaka for making them use computer facilities.

\vfill\eject
\section*{{\bf Appendix A. Splitting functions and running \\ 
       \ \ \ \ \ \ \ \ \ \ \ \ \ \ \ \ \ \ coupling constants}}
\addcontentsline{toc}{section}{\protect\numberline{\S}
{Appendix A. Splitting functions}}
\renewcommand{\theequation}{A.\arabic{equation}}

The running coupling constant in the leading order (LO) is given by
\beqn
\alpha_s^{LO}(Q^2) = \frac{4\pi}{\beta_0 \ln (Q^2/\Lambda^2)} 
\ ,
\eeqn
and the one in the next-to-leading order (NLO) is 
\beqn
\alpha_s^{NLO}(Q^2)=\frac{4\pi}{\beta_0 \ln(Q^2/\Lambda^2)}
\left[1-\frac{\beta_1 \ln \{ \ln(Q^2/\Lambda^2) \} }
{\beta_0^2\ln(Q^2/\Lambda^2)}
\right]  
\ .
\eeqn
The constants $\beta_0$ and $\beta_1$ are defined by the color constants
$C_G$, $C_F$, and $T_R$ as
\beqn
\beta_0 = \frac{11}{3} C_G-\frac{4}{3}T_R N_f  
\ , \ \ \ \ 
\beta_1 = \frac{34}{3} C^2_G - \frac{10}{3} C_G N_f - 2 C_F N_f  
\ ,
\eeqn
where
\beqn 
C_G=N_c
\ , \ \ \  C_F=\frac{N_c^2-1}{2N_c}
\ , \ \ \ T_R=\frac{1}{2}
\ ,
\eeqn
with the number of color ($N_c$=3) and the number of flavor ($N_f$).

The splitting function is expressed by the LO and NLO
splitting functions as
\beqn
\Delta_{_T} P_{q^\pm}(x) = 
\Delta_{_T} P_{qq}^{(0)}(x) + \frac{\alpha_s(Q^2)}{2\pi} \,
\Delta_{_T} P_{q^\pm}^{(1)}(x)
\ .
\label{dpij}
\eeqn
The second term is the NLO contribution. We use the expressions
in Ref. \cite{V} for the transversity splitting functions.
The LO splitting function is given by \cite{AM}
\beqn
\Delta_T P_{qq}^{(0)} (x) = C_F \Bigg[ \frac{2 \, x}{(1-x)_+} 
      + \frac{3}{2} \, \delta (1-x) \Bigg] \ ,
\label{eqn:plo}
\eeqn
where the `+' function is defined by
\beqn
\int_0^1 dz \, {\frac{f(z)}{(1-z)_+}} \ = \ 
\int_0^1 dz \, {\frac{f(z)-f(1)}{1-z}}
\ .
\label{pfun}
\eeqn
It should be noted that the above integration is defined
in the region $0 \le z \le 1$. 
However, the integration is given in the region $x \le z \le 1$
in the actual evolution calculation. 
In such a case, the integral with `+' function is calculated by
the following equation:
\beqn
\int_x^1 dz \, {\frac{f(z)}{(1-z)_+}} \ = \ 
\int_x^1 dz \, {\frac{f(z)-f(1)}{1-z}} \ + \ f(1) \ln (1-x)
\ .
\label{pfun2}
\eeqn
The NLO splitting function is \cite{V}
\begin{align} 
\Delta_T P_{q^\pm}^{(1)} (x) &\equiv \Delta_T P_{qq}^{(1)} (x) \pm 
\Delta_T P_{q\bar{q}}^{(1)} (x) \ ,
\label{eqn:pnlo1}
\\
\Delta_T P_{qq}^{(1)} (x)     &= C_F^2 \Bigg[ 1-x - \left( \frac{3}{2}
            + 2 \ln (1-x) \right) \ln x \; \delta_T P_{qq}^{(0)}(x) 
\nonumber \\
& \! \! \! \! \! \! \! \! \! \! \! \! \! \! \! \! \! \! 
  \! \! \! \! \! \! \! \! \!
                  + \left. \left( \frac{3}{8} -\frac{\pi^2}{2} + 6\zeta (3) 
                           \right) \delta (1-x) \right] 
                 + \frac{1}{2} C_F C_G \bigg [ - (1-x) 
\nonumber \\
& \! \! \! \! \! \! \! \! \! \! \! \! \! \! \! \! \! \! 
  \! \! \! \! \! \! \! \! \!
  \left.
                               + \left( \frac{67}{9} + \frac{11}{3} \ln x
                               + \ln^2 x - \frac{\pi^2}{3} \right) 
                                        \delta_T P_{qq}^{(0)}(x)  
                  + \left( \frac{17}{12} + \frac{11 \pi^2}{9} 
                  - 6 \zeta (3) \right) \delta (1-x) \right] 
\nonumber \\
& \! \! \! \! \! \! \! \! \! \! \! \! \! \! \! \! \! \! 
  \! \! \! \! \! \! \! \! \!
                               +\frac{2}{3} C_F T_R N_f  
                    \left[ \left( - \ln x -\frac{5}{3} \right) \delta_T 
           P_{qq}^{(0)}(x) - \left( \frac{1}{4} + \frac{\pi^2}{3} \right) 
           \delta (1-x) \right] \ ,
\label{eqn:pnlo2}
\\
\Delta_T P_{q\bar{q}}^{(1)} (x) &= C_F \left( C_F - \frac{1}{2} C_G \right)
\Bigg[ -(1 -x) + 2 S_2 (x) \, \delta_T P_{qq}^{(0)}(-x) \Bigg] \ , 
\label{eqn:pnlo3}
\end{align}
where $\delta_T P_{qq}^{(0)} (x) = 2x/(1-x)_+$.
The $\zeta$ function is defined by
$\zeta (k)=\sum_{n=1}^\infty 1 / n^k$,
and $\zeta(3)$ is given by the numerical value ($\zeta(3)$=1.2020569...). 
The $\pm$ factor in Eq. (\ref{eqn:pnlo1}) indicates
``$\Delta_{_T} q \pm \Delta_{_T} \bar q$ type" distribution 
$\sum_i a_i (\Delta_{_T} q_i \pm \Delta_{_T} \bar q_i)$.
The $S_2$ function is expressed in terms of the Spence function $S(x)$ as
\begin{align}
S_2(x) &\equiv \int_{x/(1+x)}^{1/(1+x)} \frac{dz}{z} \ln {\frac{1-z}{z}} 
\nonumber \\
       &= S\left({\frac{x}{1+x}}\right) - S\left({\frac{1}{1+x}}\right) 
           - \frac{1}{2} \left[\ln^2 \frac{1}{1+x}
           - \ln^2 \frac{x}{1+x} \right] 
\ ,
\label{eqn:S2}
\end{align}
where $S(x)$ is defined by 
\beqn
S(x) \, = \, \int_x^1 dz \, \frac{\ln z}{1-z} \ .
\label{eqn:SPENCE}
\eeqn
The NLO splitting function or anomalous dimensions are
the same in the $MS$ scheme \cite{KM,HKK} and
in the $\overline{MS}$ scheme \cite{V}. 
Therefore, our evolution program can be used in both schemes.

\vfill\eject

\vfill\eject
\section*{{\bf TEST RUN OUTPUT}}
\addcontentsline{toc}{section}{\protect\numberline{\S}
{TEST RUN OUTPUT}}

\noindent
IOUT= 1\ \ \ IREAD= 1\ \ \ INDIST= 1\ \ \ IORDER= 2\ \ \ 
IMORP= 2\ \ \ ILOG= 2 \\
Q02= 4.0000\ \ \ Q2= 200.000\ \ \ DLAM= 0.2310\ \ \ NF= 4\ \ \ NFI= 1 \\
XX= 0.1000000\ \ \ NX= 500\ \ \ NT=  50\ \ \ NSTEP=  50\ \ \ XMIN= 0.0001000 \\

\begin{tabbing}
     0.000100 \ \ \ \ \= 0.001041 \=  \kill
     0.000100 \>  \, 0.001041 \> \\
     0.000120 \>  \, 0.000710 \> \\
     0.000145 \>  \, 0.000292 \> \\
     0.000174 \>  --0.000226 \> \\
     0.000209 \>  --0.000860 \> \\
     0.000251 \>  --0.001628 \> \\
     0.000302 \>  --0.002547 \> \\
     0.000363 \>  --0.003638 \> \\
     0.000437 \>  --0.004921 \> \\
     0.000525 \>  --0.006418 \> \\
     0.000631 \>  --0.008149 \> \\
     0.000759 \>  --0.010137 \> \\
     0.000912 \>  --0.012398 \> \\
     0.001096 \>  --0.014949 \> \\
     0.001318 \>  --0.017798 \> \\
     0.001585 \>  --0.020945 \> \\
     0.001905 \>  --0.024379 \> \\
     0.002291 \>  --0.028073 \> \\
     0.002754 \>  --0.031979 \> \\
     0.003311 \>  --0.036023 \> \\
     0.003981 \>  --0.040097 \> \\
     0.004786 \>  --0.044055 \> \\
     0.005754 \>  --0.047704 \> \\
     0.006918 \>  --0.050799 \> \\
     0.008318 \>  --0.053040 \> \\
     0.010000 \>  --0.054072 \> \\
     0.012023 \>  --0.053492 \> \\
     0.014454 \>  --0.050863 \> \\
     0.017378 \>  --0.045742 \> \\
     0.020893 \>  --0.037724 \> \\
     0.025119 \>  --0.026499 \> \\
     0.030200 \>  --0.011925 \> \\
     0.036308 \>  \, 0.005892 \> \\
     0.043652 \>  \, 0.026521 \> \\
     0.052481 \>  \, 0.049165 \> \\ 
     0.063096 \>  \, 0.072675 \> \\
     0.075858 \>  \, 0.095662 \> \\
     0.091201 \>  \, 0.116708 \> \\
     0.109648 \>  \, 0.134650 \> \\
     0.131826 \>  \, 0.148836 \> \\
     0.158489 \>  \, 0.159176 \> \\
     0.190546 \>  \, 0.165818 \> \\
     0.229087 \>  \, 0.168430 \> \\
     0.275423 \>  \, 0.165410 \> \\
     0.331131 \>  \, 0.153773 \> \\
     0.398107 \>  \, 0.130477 \> \\
     0.478630 \>  \, 0.095291 \> \\
     0.575440 \>  \, 0.053977 \> \\
     0.691831 \>  \, 0.018859 \> \\
     0.831764 \>  \, 0.001962 \> \\
     1.000000 \>  \, 0.000000 \> \\
\end{tabbing}

\vfill\eject
\section*{{\bf Figures}}
\addcontentsline{toc}{section}{\protect\numberline{\S}{Figures}}

\vspace{2.0cm}
\begin{figure}[h]
   \begin{center}
      \epsfig{file=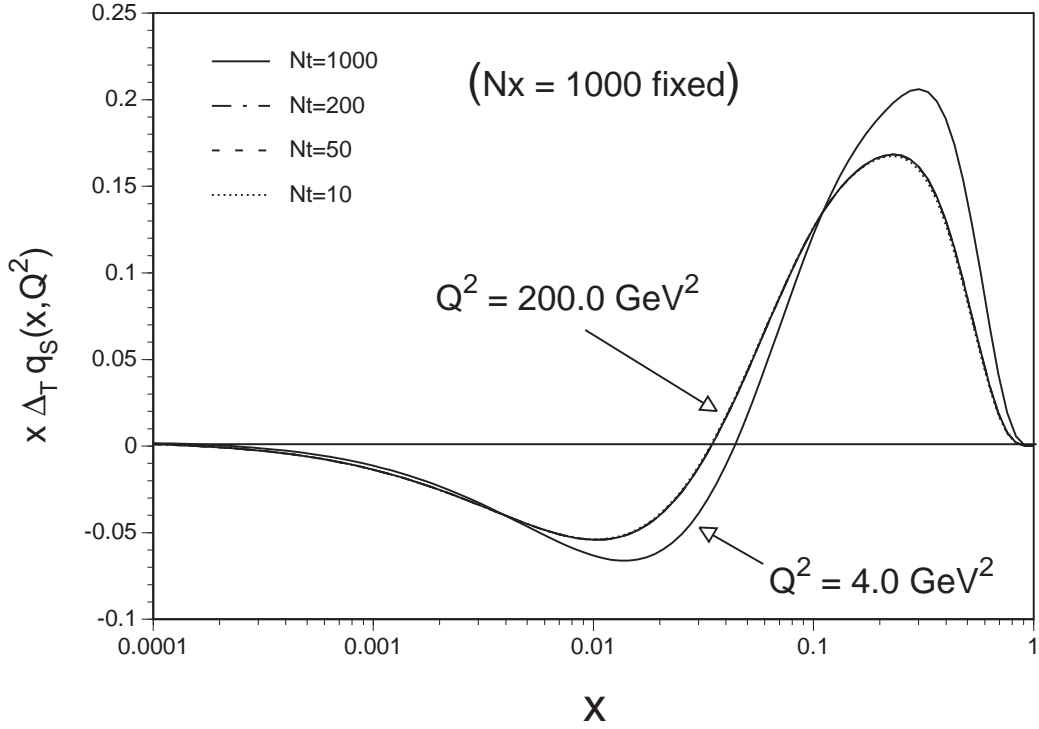,width=15.5cm}
   \end{center}
\caption{$N_t$ dependence of evolution results
is shown for the singlet transversity distribution.
The initial distribution is assumed to be the longitudinally
polarized GS-A distribution at Q$^2$=4 GeV$^2$.
It is evolved to the one at Q$^2$=200 GeV$^2$ by
the next-to-leading-order DGLAP evolution equation with
$\Lambda_{QCD}$=231 MeV and $N_f$=4.
$N_x$=1000 is fixed and $N_t$ is varied as 10, 50, 200, and 1000.
There are dotted, dashed, dot-dashed, and solid curves at 
$Q^2$=200 GeV$^2$ for $N_t$=10, 50, 200, and 1000 respectively.} 
\label{fig:nt}
\end{figure}

\vfill\eject
$\ \ \ $

\vspace{2.5cm}
\begin{figure}[h]
   \begin{center}
      \epsfig{file=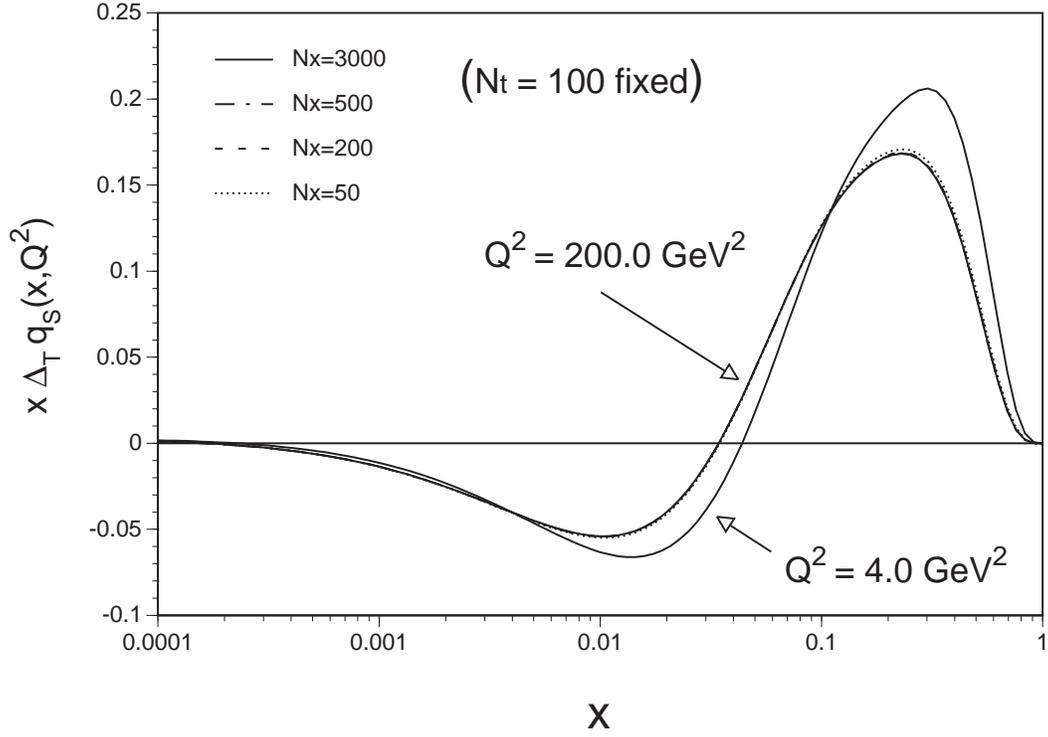,width=15.5cm}
   \end{center}
\caption{$N_x$ dependence of evolution results
is shown for the singlet transversity distribution.
The initial distribution is the same as the one in Fig. \ref{fig:nt}.
$N_t$=100 is fixed and $N_x$ is varied as 50, 200, 500, and 1000.
There are dotted, dashed, dot-dashed, and solid curves at 
$Q^2$=200 GeV$^2$ for $N_x$=50, 200, 500, and 1000 respectively.} 
\label{fig:nx}
\end{figure}

\vfill\eject
$\ \ \ $

\vspace{2.5cm}
\begin{figure}[h]
   \begin{center}
      \epsfig{file=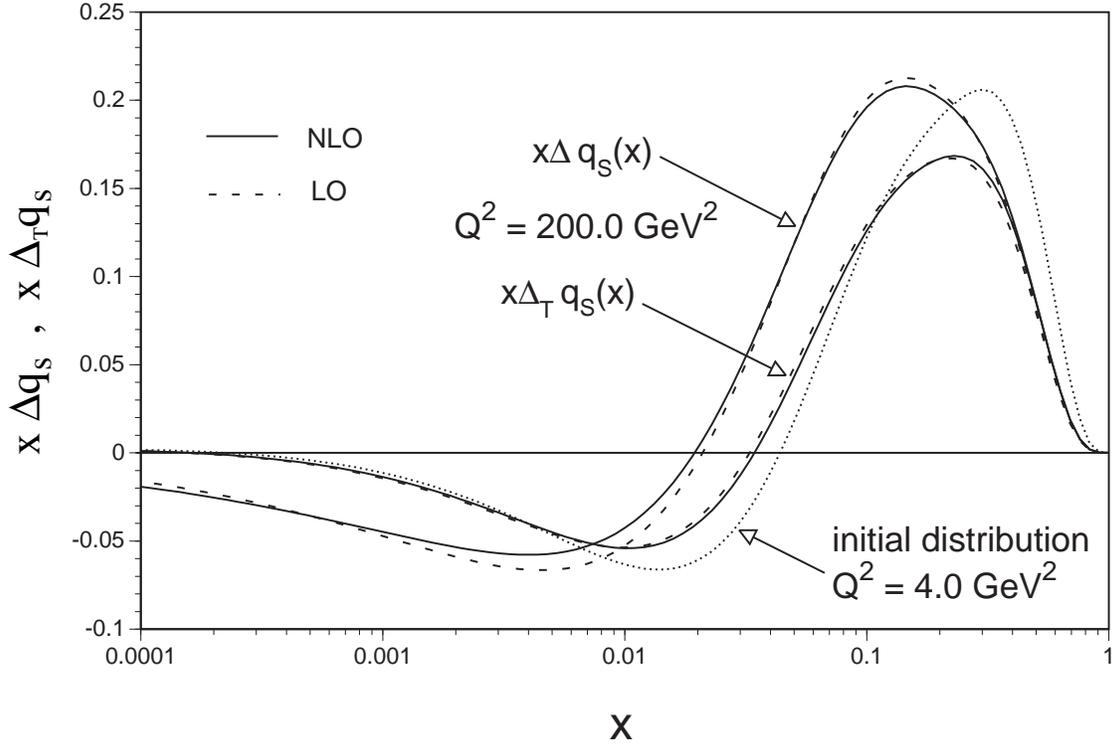,width=15.5cm}
   \end{center}
\caption{The singlet transversity evolution results
are compared with those of longitudinally polarized singlet distribution.
The longitudinal evolution is calculated by the BFP1 program \cite{BF}.
The same GS-A distribution is assumed at $Q^2$=4 GeV$^2$, and it is shown
by the dotted curve. It is evolved to the distributions at $Q^2$=200 GeV$^2$. 
The dashed and solid curves are the LO and NLO
evolution results respectively.}
\label{fig:sing}
\end{figure}

\vfill\eject
$\ \ \ $

\vspace{2.5cm}
\begin{figure}[h]
   \begin{center}
      \epsfig{file=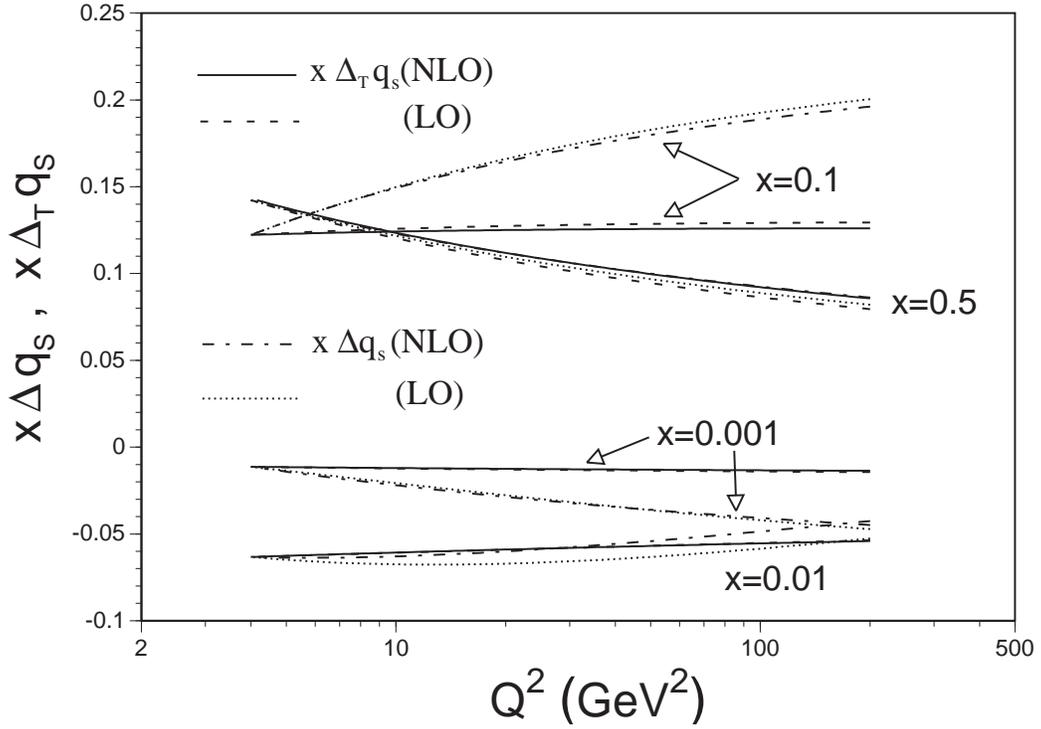,width=15.5cm}
   \end{center}
\caption{$Q^2$ dependent results of the transversity distribution are compared
         with those of the longitudinally polarized one.
         They are shown at $x$=0.001, 0.01, 0.1, and 0.5.
         The solid (dashed) curves are the NLO (LO) evolution results
         of $x\Delta_{_T} q_s$.
         The dot-dashed (dotted) curves are the NLO (LO) evolution results
         of $x\Delta q_s$.}
\label{fig:q2}
\end{figure}

\vfill\eject
$\ \ \ $

\vspace{2.5cm}
\begin{figure}[h]
   \begin{center}
      \epsfig{file=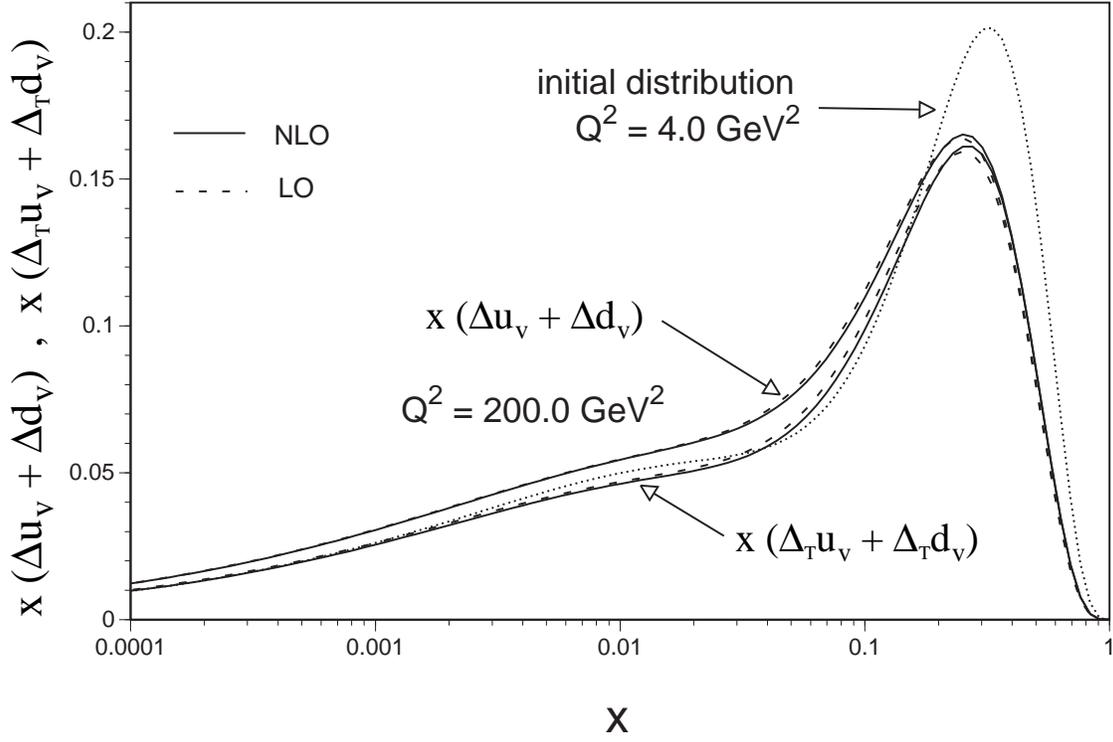,width=15.5cm}
   \end{center}
\caption{Evolution results of the valence-quark transversity distribution
$x(\Delta_{_T}u_v+\Delta_{_T}d_v)$ are compared with those of
the longitudinally polarized one. The initial GS-A distribution
at $Q^2$=4 GeV$^2$ is shown by the dotted curve, and it  
is evolved to the distributions at $Q^2$=200 GeV$^2$.
The dashed and solid curves are the LO and NLO
evolution results respectively.}
\label{fig:val}
\end{figure}

\vfill\eject
$\ \ \ $

\vspace{2.5cm}
\begin{figure}[h]
   \begin{center}
      \epsfig{file=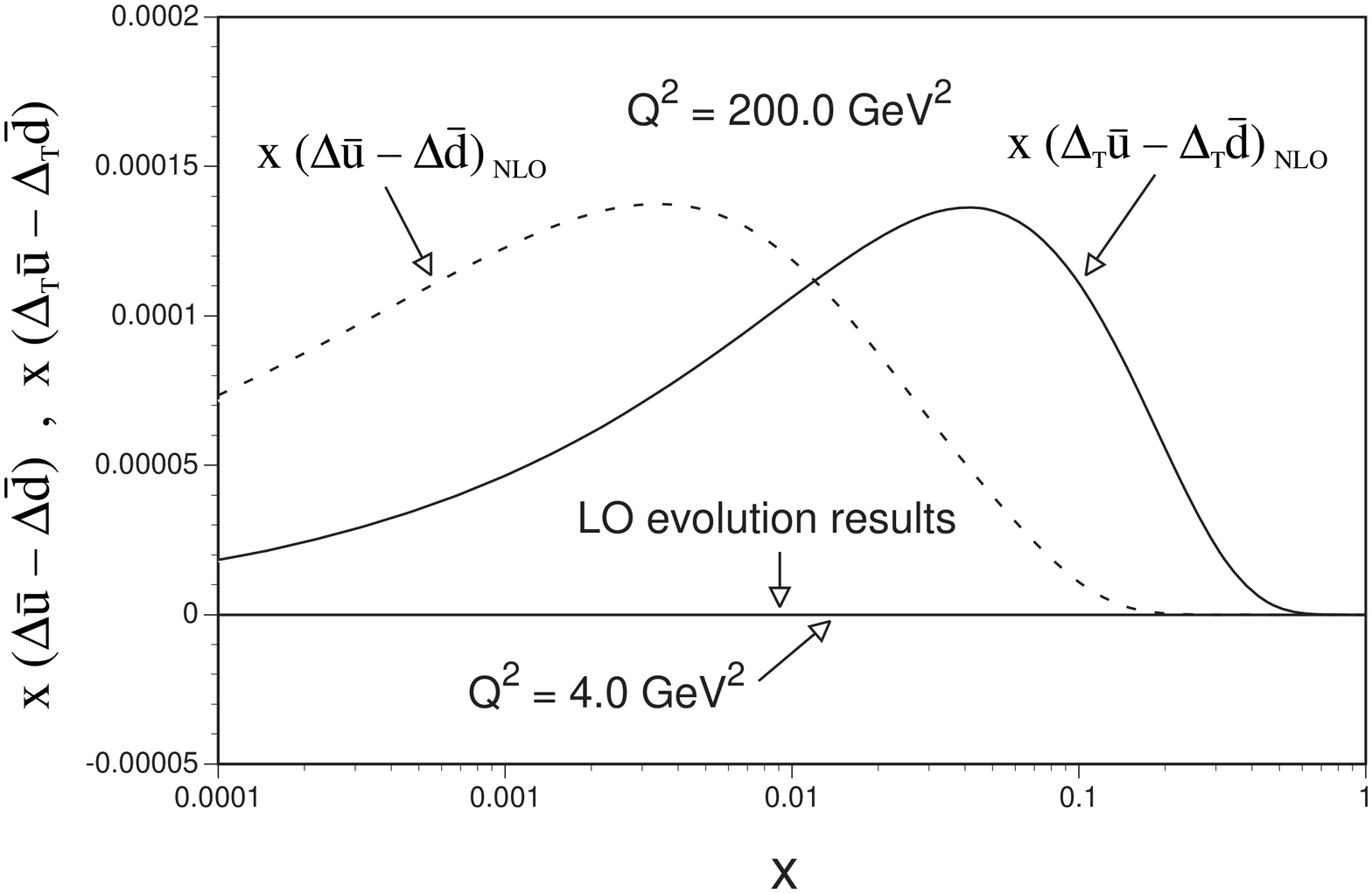,width=15.5cm}
   \end{center}
\caption{Evolution results of the flavor asymmetric transversity distribution
$x(\Delta_{_T}\bar u-\Delta_{_T}\bar d)$ are compared with those of
the longitudinally polarized one. 
There is no flavor asymmetry in the
initial GS-A distribution at $Q^2$=4 GeV$^2$ 
[$x(\Delta_{_T}\bar u-\Delta_{_T}\bar d)$=
 $x(\Delta     \bar u-\Delta     \bar d)$=0].
In the LO evolution, the evolved distributions are identically zero
[$x(\Delta_{_T}\bar u-\Delta_{_T}\bar d)_{LO}$=
 $x(\Delta     \bar u-\Delta     \bar d)_{LO}$=0].
The NLO evolution results at $Q^2$=200 GeV$^2$ are shown by 
the solid and dashed curves for the
$x(\Delta_{_T}\bar u-\Delta_{_T}\bar d)$
and $x(\Delta \bar u-\Delta \bar d)$ distributions respectively.}
\label{fig:ubdb}
\end{figure}


\begin{thebibliography}{}
\bibitem{RHIC-SPIN-J} RHIC-SPIN-J working group on parametrization:
                 Y. Goto, N. Hayashi, M. Hirai, H. Horikawa
                 S. Kumano,  M. Miyama, T. Morii, N. Saito, 
                 T.-A. Shibata, E. Taniguchi, and T. Yamanishi,
                 research in progress.
\vspace{-0.15cm}
\bibitem{RS} J. P. Ralston and D. E. Soper,
                       Nucl. Phys. B152 (1979) 109.
\vspace{-0.15cm}
\bibitem{H1}  R. L. Jaffe and X. Ji, 
                       Phys. Rev. Lett. 67 (1991) 552;
                          Nucl. Phys. B375 (1992) 527;
              X. Ji, Phys. Lett. B284 (1992) 137.
\vspace{-0.15cm}
\bibitem{DYEXP} J. L. Cortes, B. Pire, and J. P. Ralston,
                       Z. Phys. C55 (1992) 409;
                J. C. Collins, S. F. Heppelmann, and G. A. Ladinsky,
                       Nucl. Phys. B420 (1994) 565;
                B. Kamal, Phys. Rev. D53 (1996) 1142;
                R. L. Jaffe and N. Saito, 
                       Phys. Lett. B382 (1996) 165.
\vspace{-0.15cm}
\bibitem{RHIC-SPIN} Proposal on Spin Physics Using 
                   the RHIC Polarized Collider (RHIC-Spin collaboration), 
                   August 1992; update, Sept. 2, 1993.
\vspace{-0.15cm}
\bibitem{SEMI-H1} P. J. Mulders and R. D. Tangerman, 
                       Nucl. Phys. B461 (1996) 197.
\vspace{-0.15cm}
\bibitem{H1PAPER} B. L. Ioffe and A. Khodjamirian,
                       Phys. Rev. D51 (1995) 3373;
                J. Soffer, Phys. Rev. Lett. 74 (1995) 1292;
                S. Aoki, M. Doui, T. Hatsuda, and Y. Kuramashi,
                       Phys. Rev. D56 (1997) 433;
                A. Y. Umnikov, H. He, and F. C. Khanna,
                       Phys. Lett. B398 (1997) 6;
             R. Kirschner, L. Mankiewicz, A. Sch\"afer, and L. Szymanowski,
                       Z. Phys. C74 (1997) 501;
             M. Meyer-Hermann and A. Sch\"afer, hep-ph/9709349;
             K. Suzuki and T. Shigetani, hep-ph/9709394;
             R. L. Jaffe, X. Jin, and J. Tang, hep-ph/9709322.
\vspace{-0.15cm}
\bibitem{GHPRS3} M. G\"ockeler, R. Horsley, H. Perlt, P. Rakow, 
                 G. Schierholz, A. Schiller, and P. Stephenson,
                 hep-ph/9711245.
\vspace{-0.15cm}
\bibitem{AP} V. N. Gribov and L. N. Lipatov,
                       Sov. J. Nucl. Phys. 15 (1972) 438 and 675;
             G. Altarelli and G. Parisi, Nucl. Phys. B 126 (1977) 298;
             Yu. L. Dokshitzer, Sov. Phys. JETP 46 (1977) 641.
\vspace{-0.15cm}
\bibitem{AM} X. Artru and M. Mekhfi, Z. Phys. C45 (1990) 669.
\vspace{-0.15cm}
\bibitem{BCD} V. Barone, T. Calarco, and A. Drago,
                       Phys. Lett. B390 (1997) 287.
\vspace{-0.15cm}
\bibitem{KM}  S. Kumano and M. Miyama, 
                       Phys. Rev. D56 (1997) 2504.
\vspace{-0.15cm}
\bibitem{V}   W. Vogelsang, hep-ph/9706511, Phys. Rev. D in press.
\vspace{-0.15cm}
\bibitem{HKK} A. Hayashigaki, Y. Kanazawa, and Y. Koike,
                       hep-ph/9707208, Phys. Rev. D in press; 
                       numerical analysis in progress.
\vspace{-0.15cm}
\bibitem{COM} The NLO evolution was first reported in Ref. \cite{KM}
    as the Los Alamos preprint hep-ph/9706420. The first version had
    some typing and calculation mistakes. The correct final result was
    reported in the subsequent preprint hep-ph/9707208 \cite{HKK}
    within the same formalism as well as in Ref. \cite{KM}.
\vspace{-0.15cm}
\bibitem{SV}  S. Scopetta and V. Vento, hep-ph/9707250. 
\vspace{-0.15cm}
\bibitem{BST} C. Bourrely, J. Soffer, and O. V. Teryaev, 
                       hep-ph/9710224.
\vspace{-0.15cm}
\bibitem{MSSV} O. Martin, A. Sch\"afer, M. Stratmann, and W. Vogelsang,
                       hep-ph/9710300.
\vspace{-0.15cm}
\bibitem{LAG} G. P. Ramsey,
                     J. Comput. Phys. 60 (1985) 97;
              J. Bl\"umlein, G. Ingelman, M. Klein, and R. R\"uckl,
                            Z. Phys. C 45 (1990) 501;
              S. Kumano and J. T. Londergan, 
                       Comput. Phys. Commun. 69 (1992) 373;
              R. Kobayashi, M. Konuma, and S. Kumano, 
                       Comput. Phys. Commun. 86 (1995) 264.
\vspace{-0.15cm}
\bibitem{BF}  M. Miyama and S. Kumano, 
                       Comput. Phys. Commun. 94 (1996) 185;
              M. Hirai, S. Kumano, and M. Miyama, 
                       hep-ph/9707220,
                       Comput. Phys. Commun. in press.
\vspace{-0.15cm}
\bibitem{BGR} J. Bl\"umlein, B. Geyer, and D. Robaschik, hep-ph/9711405.
\vspace{-0.15cm}
\bibitem{SIMP} B. Carnahan, H. A. Luther, and J. O. Wilkes,
                      {\it Applied Numerical Methods}
                      (John Wiley \& Sons, 1969).
\vspace{-0.15cm}
\bibitem{GS} T. Gehrmann and W. J. Stirling,
                     Phys. Rev. D. 53 (1996) 6100.
\vspace{-0.15cm}
\bibitem{FMM} G. E. Forsythe, M. A. Malcolm, and C. B. Moler,
             {\it Computer Methods for Mathematical Computations}
             (Prentice-Hall, 1977).
\vspace{-0.15cm}
\bibitem{MSS} B.-Q. Ma, I. Schmidt, and J. Soffer, hep-ph/9710247.
\vspace{-0.15cm}
\bibitem{GRSV} M. Gl\"uck, E. Reya, M. Stratmann, and W. Vogelsang, 
                          Phys. Rev. D53 (1996) 4775.
\vspace{-0.15cm}
\bibitem{SK} S. Kumano, hep-ph/9702367.
\vspace{-0.15cm}
\bibitem{WASYM} N. Saito, pp.40--49 in 
             {\it Spin Structure of the Nucleon}, edited by
             T.-A. Shibata, S. Ohta, and N. Saito,
             (World Scientific, 1996);
             J. Soffer and J.-M. Virey, hep-ph/9706229;
             B. Kamal, hep-ph/9710374.
\end{thebibliography}
\end{document}